\documentclass[b5paper,twoside]{jpconf}  
\usepackage{graphicx}

\input{psfig}
\begin{document}

\title[LOFAR: status and first results ]{LOFAR: opening a new window
  on low frequency radio astronomy}

\author[Morganti et al.]{Morganti R.$^{1,2}$, Heald G.$^{1}$, Hessels J.$^{1}$, Wise
  M.$^{1,3}$,  Alexov A.$^3$, De Gasperin F.$^{4}$, Kondratiev
  V.$^{1}$, McKean J.$^{1}$, Orr\`u E.$^{5}$, Pizzo R.$^{1}$, van
  Weeren R.$^{1,6}$,  \\ on behalf of the LOFAR
  collaboration\footnote{for the full list see {\tt
      http://www.astron.nl/authors-list-lofar-commissioning-papers}}}

\address{$^1$ASTRON, Postbus 2, 7990 AA Dwingeloo, the Netherlands}
\address{$^2$Kapteyn Astronomical Institute, University of Groningen, PO Box 800,
       9700 AV Groningen, the Netherlands}
\address{$^3$       Astronomical Institute ÒAnton PannekoekÓ, University of Amsterdam, Postbus 94249, 1090 GE Amsterdam, The Netherlands}
\address{$^4$Max-Planck-Institut f\"ur Astrophysik, Karl-Schwarzschildstra{\ss}e 1,
       85741 Garching, German}
\address{$^5$ Radboud University Nijmegen, Heijendaalseweg 135, 6525 AJ Nijmegen, the Netherlands}
\address{$^6$  Leiden Observatory, Leiden University, PO Box 9513, 2300 RA Leiden, the Netherlands}

\ead{morganti@astron.nl} 

\begin{abstract}
This contribution reports on the status of LOFAR (the LOw Frequency
ARray) in its ongoing commissioning phase. The purpose is to
illustrate the progress that is being made, often on a daily basis,
and the potential of this new instrument, which is the first
``next-generation'' radio telescope.  Utilizing a novel phased-array
design, LOFAR is optimized for the largely unexplored low frequency
range: $10-240$\,MHz. The construction of LOFAR in the Netherlands is
almost complete and 8 international stations have already been
deployed as well. The wide field-of-view and multi-beam capabilities,
in combination with sub-milliJansky sensitivity at arcsec (and
sub-arcsec) resolution, are unprecedented at these frequencies.  With
the commissioning of LOFAR in full swing, we report some of the
initial results, in particular those coming from the testing of
imaging and pulsar modes.
\end{abstract}

\section{LOFAR in a nutshell}

LOFAR is a next-generation radio telescope operated by ASTRON and
constructed in the north of the Netherlands, with extensions across
Europe. Utilizing a novel phased-array design, LOFAR is optimized for
the largely unexplored low frequency range from 10 to 240\,MHz. In the
Netherlands, a total of 40 LOFAR stations are nearing completion with
an initial 8 international stations also deployed.  LOFAR has been
described elsewhere in detail (see, e.g., \cite{stappers11},\cite{haarlem12}). Here, we briefly summarize the characteristics
and the status of the telescope \footnote{For continued up-to-date
  information on the rollout of the array, the reader is referred to
  {\tt http://www.astron.nl/$\sim$heald/lofarStatusMap.html}} and of
the commissioning.

The array consists of thousands of simple dipoles, grouped into
stations.  The dipoles and stations are designed differently for the
Low-Band Array (LBA; $10-90$\,MHz) and the High-Band Array (HBA;
$110-240$\,MHz), see Figure 1.  The $90-110$-MHz range occupied by FM
radio broadcasts is filtered out.  As of early November 2011, there
are 24 core stations (within about 2\,km of the center of the array,
near the village of Exloo in the Netherlands), 9 remote stations
(within about 100\,km), and 8 international stations in France,
Germany, Sweden, and the UK (see Figure 1). An important feature of
the station design is that the HBA dipoles are split into two
substations in the core; these substations can (optionally) be
correlated separately for increased sensitivity to sources with large
angular size.
Each station is capable of forming multiple ``station beams'', which
are then correlated and/or summed as necessary in the Blue Gene P
(BG/P) supercomputer in Groningen. The product of station beams times
total bandwidth is 48\,MHz per station, giving LOFAR a remarkably
large fractional bandwidth and field-of-view (FOV).  Digital
beam-forming techniques make the LOFAR system agile and allow for
rapid repointing of the telescope as well as the potential for
multiple simultaneous observations.  The software aspect of the LOFAR
system is of crucial importance.  For example, post-processing is
handled by a number of different software pipelines which are
currently under heavy development.
 
The wide FoV and multi-beam capabilities, in combination with
sub-milliJansky sensitivity at arcsec (and sub-arcsec) resolution, are
all unprecedented capabilities at these frequencies.  LOFAR is the
most powerful and flexible low-frequency radio telescope ever
built. It is also an important precursor to the Square Kilometer Array
(SKA), by virtue of demonstrating many relevant technologies for the
first time.
 
\begin{figure}
\centerline{\psfig{file=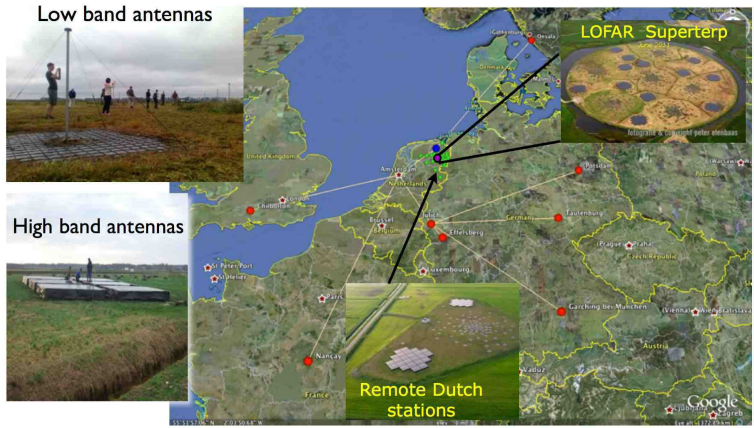,width=10cm,angle=0.}}
\caption{Map of the LOFAR stations in the {\sl ``superterp''} (inset, top
  right), remote Dutch stations (inset, bottom
  left), and international stations. On the
  left, images of the LBAs and HBAs are shown.}
\label{fig:Fig1}
\end{figure}

LOFAR will be operated as an international facility open to the
general astronomical community.  The primary scientific drivers for
LOFAR are represented by the so-called Key Science Projects (KSPs):
Surveys, Cosmic Magnetism, Epoch of Reionization, Transients, Cosmic
Rays, and Solar Science and Space Weather.  A more detailed
description of these KSPs can be found in \cite{rottgering06}.

\section{A flexible way of observing} 

A variety of observing modes are available, from standard
interferometric imaging, to tied-array beam-forming, and real-time
triggering on transients.  In fact, several of these modes can be done
in parallel, as illustrated in Figure 2.

The standard imaging mode provides interferometric data just as other
traditional synthesis arrays consisting of antenna elements. The goal
of LOFAR imaging is to achieve high fidelity and low noise images of a
range of astronomical objects, using customized observing
parameters. In this mode, station beams are transferred to the central
processing facility where they are correlated to produce raw
visibility data. The raw {\sl uv} data are stored on the temporary
storage cluster. Further processing, primarily calibration, is handled
off-line.  The pipeline which performs processing of the imaging data
is the LOFAR Standard Imaging Pipeline, which has been described by
\cite{heald10}.  The most important components of this pipeline are
(i) the flagger and data compression utility; (ii) the calibration
engine, called BlackBoard Selfcal (BBS); (iii) the imager; and (iv)
the sky model database. Flagging of radio frequency interference (RFI)
is of crucial importance. Despite the relatively high level of RFI in
northern Europe, excellent rejection without significant loss of data
is possible thanks to the high frequency and time resolution of LOFAR
data (recent observations use 4 kHz channels and $1-3$ second
integrations, depending on observing frequency). Typically, $<10$\% of
data are lost due to RFI flagging, and at many frequencies the
statistics are even better. The flagging now implemented in the
pipeline has been done using the algorithm described by \cite{offringa10}.

The imaging step itself is a difficult task for LOFAR --- the nature
of the dipoles, and their fixed orientation on the ground, makes the
sensitivity pattern of the telescope not only a function of angular
position and observing frequency, but also a strong function of
time. At the moment, LOFAR images are limited by deconvolution
errors. This is mitigated by subtracting the brightest sources in the
visibility domain prior to imaging.  The LOFAR sky model is needed for
calibrating the telescope in arbitrary locations on the sky.  An
all-sky calibration survey, aiming to produce a catalog of the
brightest sources in the LOFAR sky, has just begun (see below for
details).  Imaging capability and a summary of the sensitivity of the
array can be found in \cite{heald11}.

\begin{figure}
\centerline{\psfig{file=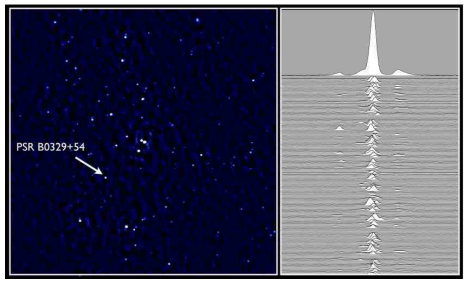,width=6cm,angle=0.}}
\caption{LOFAR HBA observations of the pulsar B0329+54 simultaneously
  using both the imaging (left) and beam-formed (right) modes. The
  location of pulsar is indicated by the arrow.}
\label{fig:Fig2}
\end{figure}

There are many ways in which the LOFAR antennas and stations can be
combined to form beams relevant for observing pulsars
and fast transients.  A detailed description is given in
\cite{stappers11}.  In short, the term station beam corresponds to the
beam formed by the sum of all of the elements of a station. For any
given observation there may be more than one station beam and they can
be pointed at any location within the wider element beam. A
tied-array beam is formed by coherently combining all the station
beams, one for each station, which are looking in a particular
direction.  To sample the combined radio signal at significantly
higher time resolution (t$_{\rm samp} < 100$ ms), one has to normally
sacrifice spatial resolution, and/or the large FoV seen by the
individual elements, to form a single beam pointing in the direction
of the source of interest.  To compensate for this, there may be more
than one tied-array beam for each station beam. Station beams can also
be combined incoherently in order to form incoherent array
beams. These retain the FoV of the individual station beams but have
increased sensitivity compared with a single station.  A pulsar
pipeline has been developed to cope with the data and a description
can be found in \cite{alexov10}.
Finally, the LOFAR Transients Key Science Project (\cite{fender06})
will use both the imaging and beam-formed modes to discover and study
transient sources. The imaging mode will probe flux changes on
timescales of seconds to years, while the beam-formed modes will probe
timescales from seconds down to microseconds. With the Transient
Buffer Boards it will be possible also to form images on very short timescales.

\section{Coping with the data rate and volume} 

LOFAR is in the vanguard of new astronomical facilities dealing with
the transport, processing, and storage of extremely large amounts of
data.  The raw data-rate generated at station level by the entire
LOFAR array is 13\,Tbit/s, far too much to transport in its
entirety. Even though this raw data-rate is reduced by, e.g.,
beam-forming at station level, the long range data transport rates
over the array are still of order 150\,Gbit/s, requiring partially
dedicated fibre networks. Such large data transport rates naturally
also imply data storage challenges. For example, typical
interferometric imaging observations can easily produce 35 TBytes/hr
of raw, correlated visibilities.
 
The LOFAR station data are sent via a high-speed (partly dedicated)
fiber network infrastructure to a central processing facility located
in Groningen, in the north of the Netherlands. At this central
processing (CEP) facility, data from all stations is aligned in time,
combined, and further processed using a Blue Gene/P supercomputer
offering about 28\,TFLOP of processing power. As mentioned above, the
Blue Gene/P performs a variety of processing operations, including
correlation for standard interferometric imaging, tied-array
beam-forming for high time resolution observations, and even real-time
triggering on incoming station data streams. Combinations of these
operations can also be run in parallel. Blue Gene/P writes raw data products to a storage cluster for
additional post-processing. At the moment, storage limits give a $\sim
1$ week processing window. When complete, this cluster will host
2\,Pbyte of working storage. As mentioned above, once on the storage
cluster, a variety of reduction pipelines are then used to further
process the data into the relevant scientific data products depending
on the specific type of observation. Science-specific pipelines run on
a dedicated compute cluster with a total processing power of
approximately 10\,TFLOP.

\begin{figure}
\centerline{\psfig{file=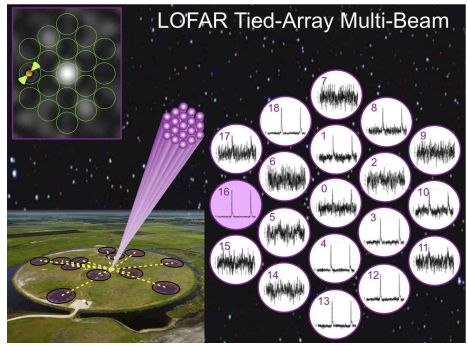,width=6cm,angle=0.}}

\caption{Test observation of the tied-array multi-beam mode, in which
  19 beams have been synthesized simultaneously. The pulsar B0405+55
  is located in beam 16.}

\label{fig:Fig3}
\end{figure}

\section{Commissioning and first results}

As commissioning continues, the first science results from LOFAR are
beginning to appear. Many of these results were presented at the LOFAR
Workshop {\sl ``First Science with LOFAR''} and they can be found at
http://www.astron.nl/lofarscience2011/ .

For pulsars, the LOFAR frequency range is an under-explored part of
spectrum.  Observations of known pulsars have already brought more
than 100 detections and preliminary results, as well as LOFAR's pulsar
observing modes are described in detail in Stappers et al. (2011).  A
coherent sum of multiple stations is now routinely used to perform
high-sensitivity, high-time-resolution observations of, e.g., pulsars,
(exo)planets, flare stars, and cosmic rays.  Furthermore, to
compensate for the reduced FoV of these tied-array beams, multiple
simultaneous beams are now regularly used (see Figure 3).  Recently
the pulsar group has started testing LOFAR observations that use 127
tied-array beams synthesized from the Superterp stations.  This mode
provides an excellent tool for sensitive large-area surveys. Indeed, a
large pilot survey for pulsars has been taken, using a coherent
{\sl "superterp"} and 19 tied-array beams, providing 3.7 deg$^2$ of sky per
pointing (Coenen et al. in prep).

The low-frequency range and large fractional bandwidth of LOFAR
provide a unique view of the pulsar emission process.  LOFAR data is
already providing interesting insight into the frequency evolution of
pulse profile morphology, placing constraints on the emission height
within the pulsar's magnetosphere (Hassall et al. in prep.).  LBA
observations of pulsars B0809+74 and B1133+16 show very bright,
sporadic single-pulses, which also provide insight into the emission
mechanism (Kondratiev et al. in prep.).

\begin{figure}
\centering
\centerline{\psfig{file=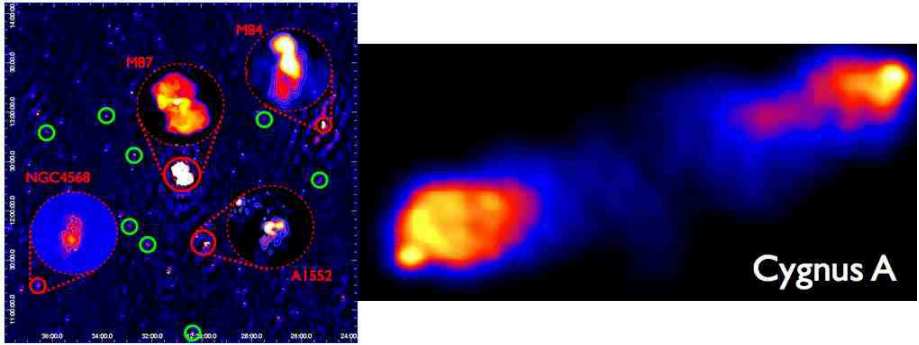,width=12cm,angle=0.}}
\caption{LOFAR images of Virgo~A and the surrounding sources at 135 MHz (left) and Cygnus~A
  (right). Good models of these very strong sources are important for
  LBA imaging, as they need to be subtracted from the visibilities
  regardless of the pointing direction.}
\end{figure}
 

The commissioning period is also producing some excellent imaging
results (see e.g. \cite{heald10} and \cite{mckean11}), in spite of
various complications due to the low frequency and wide FoV.  For
instance, the LBAs have an extremely large FoV, and consequently the brightest
sources (e.g. Cassiopeia~A, Cygnus~A, Virgo~A and Taurus~A; the
so-called ``A-team'')  in the sky, which are omnipresent in the side-lobes,
must be subtracted in order to properly calibrate the data (e.g. beam changing over
time, ionosphere).  Brute-force strategies, solving for the
gains in the directions of the brightest sources and then subtracting
them from the visibilities, have been successful but are highly
computationally expensive. Recently, the ``demixing'' method described by
\cite{vdtol07} has been tested and found
remarkably successful. This method requires a good model of the A-team
sources and, therefore, some of the highest-quality images made so far
are of, e.g., Virgo~A and Cygnus~A (see Fig. 4). This method has
shown that not only the target source can be well calibrated and
imaged following demixing, but also the off-axis bright sources as
well --- in one recent test, Cassiopeia A was successfully imaged,
despite being located some $127^{\circ}$  away from the pointing center.

The detailed study of these radio sources has also provided initial
results on their spectral indices. The radio spectrum of Cygnus A has
been traced from 240~MHz down to 30~MHz (McKean et al. in prep). A
steepening of the spectral index towards the centre is observed, as
expected for the synchrotron aging model. Similar spectral indices are
found in the two lobes.  In the case of Virgo~A, the new observations
also demonstrate the impressive wide-field capabilities of LOFAR,
which allow simultaneous imaging of the other sources in the cluster
(see Fig. 4 left; De Gasperin et al. in prep).  Figure 5
illustrates other example images from commissioning observations.  The
topics that are currently receiving the most attention are the study
of normal galaxies (illustrated by NGC 4631 where the halo of radio
emission is starting to become visible; Jurusik et al. in prep), giant
radio galaxies (e.g. B1834+620, where also polarized emission has been
detected; Orr\'u et al. in prep.)  and cluster halos (e.g. A2256; van
Weeren et al. in prep.).

Given the importance of the initial sky model for the calibration of
LOFAR data, a survey aimed at improving this model has just
started. The "Multifrequency Snapshot Sky Survey (MSSS) " (Heald et al. in prep) will be done
both at high-band as well as low-band frequencies and it will reach
respective noise levels of $<15$ and $<5$ mJy/beam. However, the spatial resolution will be 
relatively low  ($\sim 100$ arcsec) as it will make use only
of the shorter baselines (core stations) due to limited computing
power. Visibility data from longer baselines will be kept available for possible post-processing that will result in higher resolution images. An example of the large fields that will be imaged in one
pointing is given in Fig. 6.

\begin{figure}
\centering
\centerline{\psfig{file=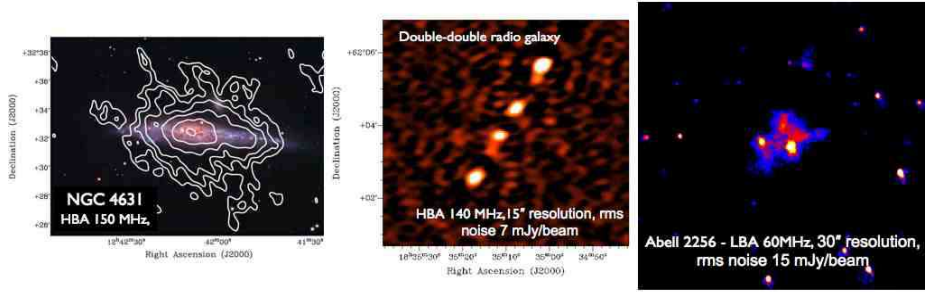,width=12cm,angle=0.}}

\caption{Example of initial results from LOFAR commissioning
  observations relevant for the study of normal galaxies (NGC4631),
  giant radio sources, and clusters halos.}

\end{figure}

Finally, a few words about two other important KSPs. The Epoch of
Reionisation (EoR) is the most challenging project that LOFAR will
carry out. The efforts of the EoR group are now concentrated on the
correction of direction-dependent effects, and more specifically
imaging artifacts due to the different and, to some degree uncertain,
station beams. Using the new SAGECAL (\cite{kazemi11}) software
package, calibration in up to 100 directions is now possible within 14
hours of processing. This has made it possible to approach the thermal
noise, even though a proper ionospheric calibration is not yet in
place. However, this situation might change in the coming months/years
as the solar activity increases. Another milestone is the pipeline
processing of wide FoV observations using multiple LOFAR
beams. Finally, there is an on-going effort on increasing the
performance and accuracy of calibration and imaging algorithms on
multi-core architectures.

\begin{figure}
\centering
\centerline{\psfig{file=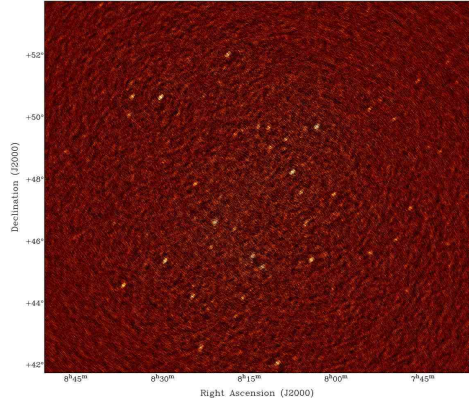,width=6cm,angle=0.}}
\caption{LOFAR LBA image of the 3C196 field (the strong central
  source has been removed). Note the size of the image, which
  covers about 100 $\opensquare^2$.}

\end{figure}

The goal of the Cosmic Ray KSP is to detect the radiation produced
when a cosmic ray hits the Earth's atmosphere, producing a cascade of
secondary particles (see \cite{corstanje11}).  This is used to study
the properties of the primary cosmic rays as well as the development
of the particle cascade and the radiation mechanism. LOFAR can trigger
on cosmic rays in two ways: directly on the radio air shower signal,
or via an external trigger (e.g. from a particle detector array). At
present, the detection of cosmic rays by LOFAR has been done using a
particle detector array as trigger: the Lofar Radboud Air shower array
(LORA). When LORA detects a cosmic ray signal it requests a read-out
of a ring-buffer (Transient Buffer Board data) from each LOFAR dipole
on the {\sl "superterp"}.

In summary, LOFAR is almost fully constructed, the commissioning is in
full swing, and the first science results are beginning to
appear. Challenges, especially in the calibration and imaging, still
need to be solved but also this is proceeding very fast. With its
dense core array and long interferometric baselines, LOFAR is about to
change our view of the low-frequency Universe!

\section*{Acknowledgments}
LOFAR, the Low Frequency Array designed and constructed by ASTRON, has
facilities in several countries, that are owned by various parties
(each with their own funding sources), and that are collectively
operated by the International LOFAR Telescope (ILT) foundation under a
joint scientific policy.  The results presented here are the
culmination of a great deal of work done by a large number of
people. We gratefully acknowledge the engineers who have designed and
built the array, as well as the large group of commissioners who have
taken part in LOFAR ``Busy Weeks,'' where the bulk of the
commissioning progress has taken place.

\end{document}